\documentclass[referee,a4paper,12pt,traditabstract]{swsc} 

\usepackage{graphicx}
\usepackage{txfonts}
\usepackage{subfigure}
\usepackage{epstopdf}
\usepackage{lineno}
\usepackage[backref]{hyperref}
\usepackage{url}

\def\ddt#1{{{\partial #1}\over {\partial t}}}

\def\solphys{{\it Solar Phys.}}
\def\ApJ{{\it Astrophys. J.}}
\def\apjs{{\it Astrophys. J. Sup.}}
\def\aap{{\it Astron. \& Astrophys.}}

\hypersetup{colorlinks=true,citecolor=cyan,urlcolor=cyan,linkcolor=blue}

\begin{document}


   \title{Automated Detection of Solar Eruptions}

   \titlerunning{Solar Eruptions}

   \authorrunning{Hurlburt}

   \author{N. Hurlburt }

   \institute{Lockheed Martin Solar and Astrophysics Laboratory\\
              Lockheed Martin Advanced Technology Center\\
              3251 Hanover Street, Palo Alto, CA, USA}

\date{Received November 21, 2014; accepted December 1, 2015}

 
  \abstract
   {Observation of the solar atmosphere reveals a wide range of motions, from small scale jets and spicules to global-scale coronal mass ejections. Identifying and characterizing these motions are essential to advancing our understanding of the drivers of space weather. Both automated and visual identifications are currently used in identifying Coronal Mass Ejections (CMEs). To date, eruptions near the solar surface, which may be  precursors to CMEs, have been identified primarily by visual inspection. Here we report on Eruption Patrol (EP): a software module that is designed to automatically identify eruptions from data collected by SDO/AIA. We describe the method underlying the module and compare its results to previous identifications found in the Heliophysics Event Knowledgebase. EP identifies eruptions events that are consistent with those found by human annotations, but in a significantly more consistent and quantitative manner. Eruptions are found to be distributed within 15 Mm of the solar surface. They possess peak speeds ranging from 4 to 100 km/s  and display a power-law probability distribution over that range. These characteristics are consistent with previous observations of prominences.
    }        


   \maketitle

\section{Introduction}

 Eruptions in the low solar atmosphere are key elements in generating space weather. Large eruptions can evolve into coronal mass ejections (CMEs) that can plow through the solar wind and ultimately impact the earth's magnetosphere (\cite{Munro}, \cite{Gosling}). The source of many of these CMEs has been associated with prominence eruptions, e.g., \cite{Gopal2003}.  \cite{Yan2011} find that approximately half of active region filament eruptions are associated with CMEs and over 90\% are associated with flares. Smaller eruptions may provide the ultimate source for the solar wind (\cite{Tian}). Regardless of their magnitude, eruptions play a significant role in the structure and dynamics of the solar atmosphere.
  
Identifying eruptions occurring near the solar surface is complicated by the presence of a wide variety of features and scales. Active regions, coronal holes and filaments persist for long periods, while the short-lived eruptions pass through and among them. Flares and other sudden changes in intensity add distractions that can mask or mimic motions that would otherwise be visible. As a result, automated detection of these eruptions has been challenging.

Previous studies have developed automated methods to detect and track filaments primarily in H-alpha images (\cite{Gao}, 
\cite{2005SoPh..228...97B}, \cite{2005SoPh..228..137Z}, \cite{2005SoPh..228..119Q}, \cite{2008AnGeo..26..243A}, \cite{2008SoPh..248..425S}, \cite{2010SoPh..262..425J}, \cite{2011SoPh..272..101Y}, \cite{2013SoPh..286..385H}, \cite{Schuh}). These methods typically use image-based feature detection followed by a tracking step comparing the results of sequential detections. Measured velocities for prominence eruptions in these studies tend to lie in the range of 10-100 km/s  while quiescent prominences show velocities around 4 km/s or less.

We take a different approach by first extracting velocities from a sequence of images and then identifying features within the resulting velocity fields. Here we use an optical flow method to identify regions of significant motion. The use of these velocity fields to define regions of interest, rather than working directly from the images, removes many of the distracting features and permits us to identify and  characterize the flows in sufficient detail for further analysis. We note that \cite{Gissot2008} have previously applied an optical flow method to an erupting filament using simultaneous pairs of 30.4nm images acquired by the EUVI instruments on the two STEREO spacecraft (\cite{Kaiser2008}). However their objective was to reconstruct the three-dimensional structure of a filament that had been identified by visual inspection rather than to find the eruption autonomously.

In the following sections we present the underlying method used by Eruption Patrol (EP), assess its performance, survey the statistical properties of the resulting detections and summarize our findings in the conclusion.

\section{Method}
Our approach to identifying solar eruptions is to extract  velocity fields from sequences of solar images using the \textsf{opflow3d} method described in \cite{HJ} as applied to images obtained by the Atmospheric Imaging Assembly on the Solar Dynamics Observatory (SDO/AIA, \cite{Lemen}).  The Interactive Data Language (IDL) routine, \textsf{opflow3d}, creates an estimate for an unknown, steady, velocity field ${\bf v}(x,y)$ from a time sequence of images $I(x,y,t)$ with the assumption that it satisfies the advection equation
\begin{equation}
\ddt{I}+ {\bf v} \cdot \nabla I = 0. \label{eq:advect}
\end{equation}
The best fit velocity field ${\bf v}_f$ can then be found using least squares applied over
a set of images $I(x,y,t)$ over some time interval, $\Delta t$. \textsf{opflow3d} expands the fit velocity ${\bf v_f}$ in a truncated Fourier series
\begin{equation}
{\bf v}_f=  \Sigma_{i=-N_x}^{i=N_x}\Sigma_{j=-N_y}^{j=N_y}(\alpha_{ij}{\bf\hat x} +
\beta_{ij}{\bf\hat y}) e^{-2\pi{J( ix/X + jy/Y)}}, \label{eq:vel}
\end{equation}
where $J=\sqrt{-1}$, and $X$ and $Y$ are the spatial dimensions of the image set. It then determines the optimal values of the complex amplitudes $\alpha_{ij}$
and $\beta_{ij}$, where {$\bf\hat x$} and {$\bf\hat y$} are unit vectors, and $ N_x$ and $N_y$ are the
number of Fourier modes retained in the expansion. 

The primary goal of EP is to identify regions of high velocity while operating within the near-real time AIA data pipeline. Hence we choose parameters for EP that can keep up with the data flow. By setting $\Delta t=2$ min, $N_x, N_y =8$ and $X,Y=4096$ pixels, \textsf{opflow3d} can extract a velocity fit from the full-resolution AIA images in approximately one minute of computation on a 2013-vintage Apple iMac.  The effective resolution element of this fit is one quarter of the shortest wavelength captured in equation \ref{eq:vel}, which is $0.25 X/N_x=128$ pixels, or about 80 arcsec.  
Systematic errors are generally negligible given the quality and consistency of AIA data, but there may be some ringing due to Gibbs phenomena when strong motions are present at the edge of the field of view. Additional sources of error are discussed below.

A two-minute sequence of ten, full-resolution Level 1 images taken in the He II line at 30.4 nm is processed to create one velocity map. These images have had dark current and flat-field corrections and spikes caused by bad pixels and radiation hits removed. The intensity in AIA images is expected to scale as the square of the density of the emitting plasma multiplied by a function temperature (\cite{1983ApJS...52..155G}). Over the two minutes used in determining the fit, we expect that changes in intensity are mostly due to changes in density since temperature changes are dampened relatively quickly  (\cite{1982soma.book.....P}) except in actively flaring regions. Hence, the square-root of these images is a proxy for the density, which we expect to approximately obey equation \ref{eq:advect}. This is consistent with the visual impression given by viewing the image sequences. We therefore apply a square-root to the images before passing them into \textsf{opflow3d}. 

Previous studies found velocities exceeding 100 km/s, which translates to about 17 arcsec over the two minute sampling period used for a velocity fit. This is well within our effective resolution so systematic error due to smearing should be small; it also suggests that the maximum reliable velocity estimate we can expect for our sampling choice is about 500 km/s, i.e., a displacement of 80 arcsec over the interval $\Delta t=2$ min. Fitting for higher speeds would require either smaller $\Delta t$ or larger spatial windows. For instance, we could detect speeds approaching 5 Mm/s over 80 arcsec by setting $\Delta t = 12$s, the maximum AIA cadence. The minimum speed of the \textsf{opflow3d} method itself is only limited by roundoff errors. Note that if we were to apply this method to the coronal images collected by AIA, say 19.3 nm where velocities are expected to reach these ranges, we would need to adjust our parameters accordingly. In that case, the more complex variations in intensity may require an explicit treatment, such as the multi-scale optical flow method described by \cite{2007A&A...464.1107G} which was used by \cite{Gissot2008}. 

The derived velocity fields are composed of multiple components, some of which are sources of error for our application: these include solar rotation, super-granulation and other quasi-static motions. Most of these motions are small and reasonably isotropic. The solar rotation profile is neither, with a peak value of about 2.2 km/s. This can introduce a bias when using a thresholding technique to identify eruption sites. Hence, EP subtracts a background velocity corresponding to that of solid-body rotation.  We do not attempt to remove smaller effects such as differential rotation and meridional circulations. 

\begin{figure*}
   \centering
   \includegraphics[width=\columnwidth]{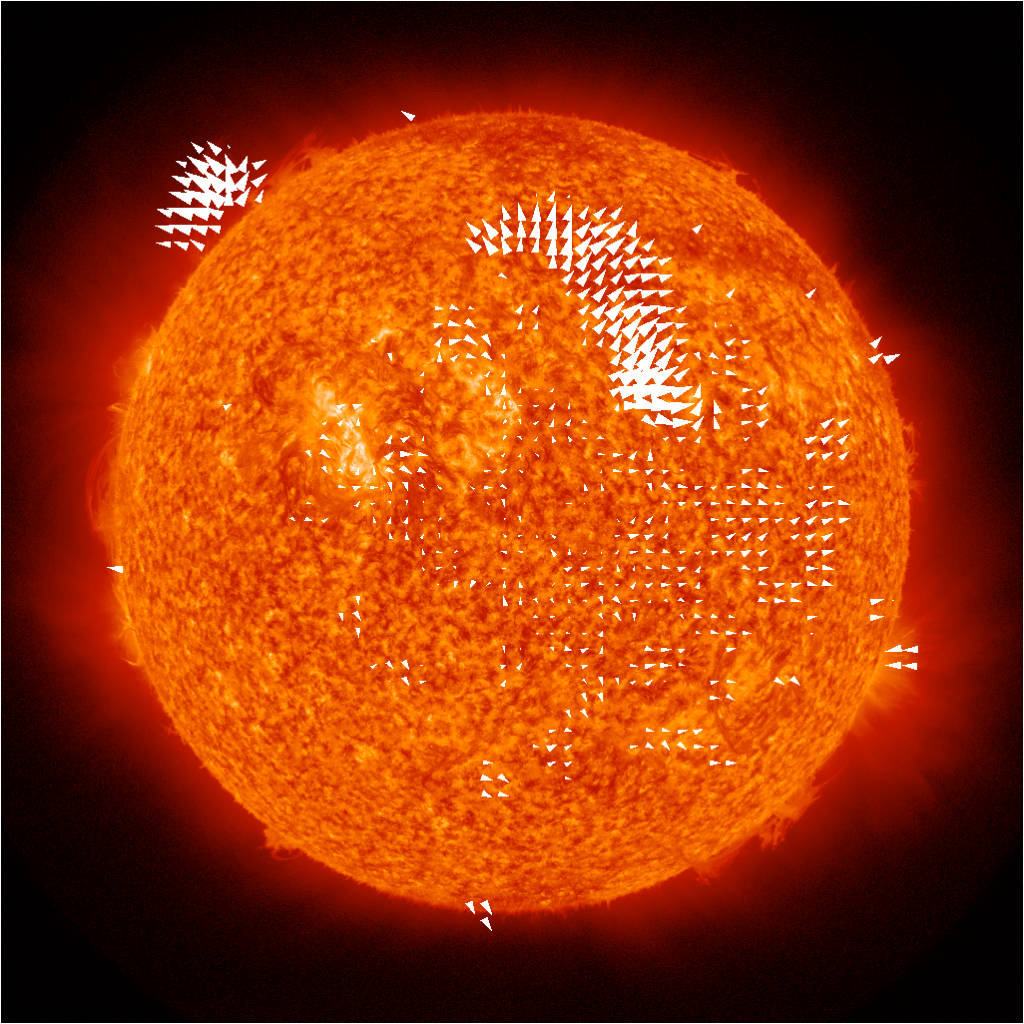}
   \caption{\small The velocity field obtained by applying our method to two minutes of SDO/AIA 30.4 nm data starting at 2010-Aug-01T21:20. The arrows are aligned with the local velocity with areas proportional to the speed.  The corresponding image is shown in the color background. Two regions are seen to be erupting: a long filament on disk is ascending into the corona; and another region in the upper left that may be part of the same eruption, or a sympathetic response. Only velocities over 1 km/s are displayed and solar rotation is not removed. The peak velocity here is 3.8 km/s.} 
   \label{fig:Aug2010}
\end{figure*}

An example of the resulting flow field is displayed in Figure \ref{fig:Aug2010}. The spatial resolution of our velocity fit was doubled to 30 arcsec here to better define the regions. The flow associated with the large filament eruption near the northern pole is clearly captured, as well as a few smaller-scale flows around the limb.

\begin{figure}
   \includegraphics[width=\columnwidth]{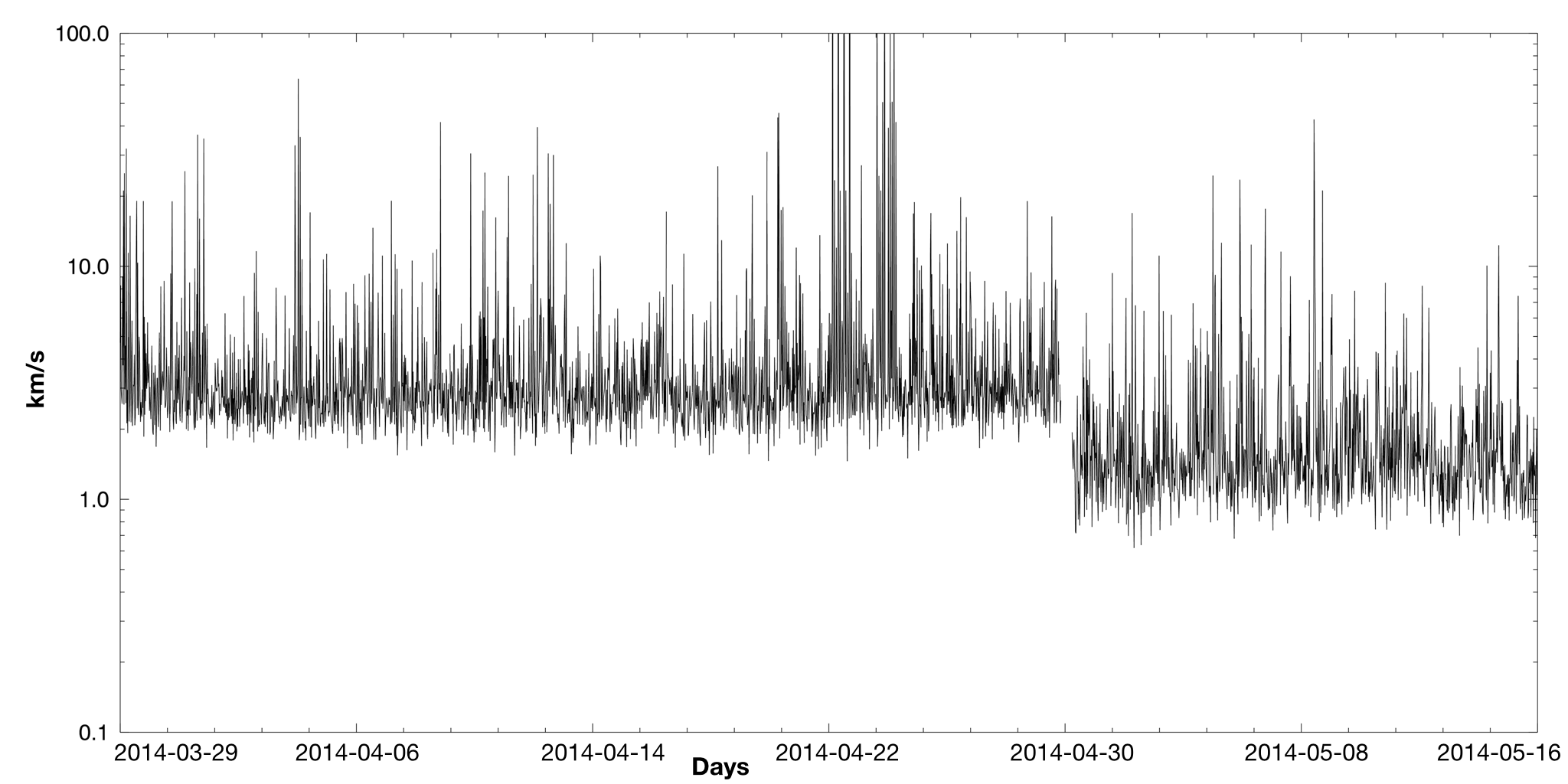}
   \caption{\small The peak speeds reported by EP in 3000 samples taken between 2014-03-29 to 2014-05-16. The preliminary results for April did not remove the effects of solar rotation, and hence have a floor of about 2 km/s. Data in May have that correction applied, and the remaining floor is a combination of differential rotation, meridional flow, super-granulation and other ubiquitous sources. Spacecraft maneuvers and other calibrations are also present in the April results, most notably during April 23.} 
   \label{fig:raw}
   \end{figure}

EP samples velocities every 20 minutes and records the time, location and velocity at the point of maximum speed within each sample. As described above, this velocity corresponds to the best-fit over a region of about 80 arcsec in a two minute interval. Hence the precise position of the peak is only known to that resolution. Figure \ref{fig:raw} displays the raw output of the patrol over a seven week period starting on 29 March 2014. The effect of the rotation removal can be seen as a drop in the floor of the velocity measurements to values consistent with those expected from super-granulation (e.g., \cite{Shine}) and other sources. Peaks corresponding to eruptions and spacecraft motions are also clearly visible. The latter are excluded in the production version of the method.

The decision to only sample 10\%  of the images, i.e., 2 out of 20 minutes, risks missing short-lived events. However eruptions with lifetimes shorter than 20 minutes probably have little impact on their surroundings. Our goal here is to identify eruptions that may have significant impacts, so the computational cost savings outweigh the loss of information. Our intent is to return to periods of significant motion for more detailed analysis in the future.

The results of this first pass are then processed to identify time periods where velocity exceeds the threshold of 3.6 km/s. This value was settled upon by the need to exclude the background motions seen in Figure \ref{fig:raw} while generating a moderate detection rate. It also corresponds to the level of motion found by \cite{Wang} in quiescent prominences. These periods, along with the largest velocity and its position, are then recorded to the Heliophysics Events Knowledgebase (HEK, \cite{Hurlburt}) as preliminary reports of eruptions. Our intent is to analyze these more carefully in a second "characterization" routine and then replace or update these entries with more details.

\section{Results}
\subsection{Comparison with manual selection}
We assess the performance of our method by comparing it to eruptions recorded manually to the HEK. These entries are primarily provided by members of the SDO/AIA science team who monitor data as they arrive at the AIA Validation Center (see \cite{Hurlburt} for details).  The volunteer annotators regularly sign up for three-day shifts. Thus all the datasets used by EP, along with other AIA channels, have also been reviewed by this team. Over the interval from 18 April 2014 to 17 July 2014, a total of 43 filament eruptions and 44 eruptions were recorded by the team. For this case we consider an eruption to be any of two classes accepted by the HEK: eruptions and filament eruptions. The first is a catch-all category that may or may not be associated with a filament; the later is associated with a filament that the observer considered to have ejected material into the corona. 

As a first test, we queried the HEK for both classes of eruptions using iSolsearch\footnote{\url{http://www.lmsal.com/isolsearch}} to select the events and then exporting them into SolarSoft (\cite{Freeland}) and using the IDL routine \textsf{hek\_match\_events}. For this study we considered events that overlapped within an hour in time. The results are displayed in Table 1. Of the 43 filament eruptions reported by humans, 37 (79\%) matched times reported by EP. The success drops to 24 (44\%) when we also require a separation of less than 120 arcsec. Human reports of eruptions displayed a similar behavior, which is partly due to some observers selecting both when they generate their reports.  We will discuss this further below.

\begin{table*}
\label{table:1}      
\centering                          
\begin{tabular}{c c c c c c}        
\hline\hline                 
Comparison & Count& Hit (t) & Miss (t) & Hit (x,t) & Miss (x,t) \\    
\hline                        
EP from FE  & 43 & 79\% & 21\% &44\%&56\% \\      
EP from  ER & 44& 84\%    & 16\%&55\%&45\% \\
FE+ER from EP & 29 & 31\% & 69\%&24\%&86\% \\
 
\hline                                   
\end{tabular}
\caption{A bi-directional comparison between filament eruptions (FE) and generic eruptions (ER) reported manually and EP reports over the  time span from 4/18/2014 to 7/17/2014. The first two rows show the success of EP in matching times recorded by FE and ER. The third row shows the success of the manual reports in matching times reported by EP. The columns show the total number of reports in each set, the percentage of hits and misses for the reported times, and hits and misses for both time and position.}   
\end{table*}
\begin{table*}
\label{table:2}      
\centering                          
\begin{tabular}{c c c c c c}        
\hline\hline                 
Comparison & Count& Hit (t) & Miss (t) & Hit (x,t) & Miss (x,t) \\    
\hline                        
 EP from FE  & 813 & 68\% & 32\% &27\%&73\% \\      
EP from ER& 327& 70\%    & 30\%&53\%&47\% \\
FE+ER from EP& 289 & 24\% & 76\%&11\%&89\% \\
 
\hline                                   
\end{tabular}
\caption{A similar comparison as in Table 1, but over the large time span from 5/15/2010 to 11/17/2014.}
\end{table*}

As a second test, we selected the 29 events with speeds exceeding 30 km/s from EP over the same interval and compared them to entries reported by human annotators.  Nine (31\%) match the human annotations in time, while the remaining 20 did not. Only 7 of those 9 also overlapped spatially. All of these missed events were reviewed visually using the daily movies posted by the AIA team\footnote{\url{http://sdowww.lmsal.com}} and were found to be associated with significant eruptions. 

These patterns persist over the entire AIA dataset, as can be seen in Table 2. Overall EP finds about 70\% of all time periods manually reported as erupting in either category.  This moderate drop for Hit(t) in Table 2, as compared to Table 1, may be due to the fact that  humans are particularly good at catching slow, long-duration eruptions. Comparisons of EP detections exceeding 30 km/s to manual ones shows similar drop from 31\% in Table 1 to 24\% in Table 2. This might indicate that the converse is also true: fast, short-duration eruptions may be overlooked by human annotators. This is consistent with what we found in the smaller sample, but we have not manually reviewed the 289 events to confirm this.

The success at matching both the time and location of filament eruptions drops to 27\% in the larger sample. This may again be due to large-scale filament eruptions. In that case, the position selected by annotators will tend to be the geometric center of the filament, while the position reported by EP will be the fastest moving element, or perhaps that of a separate fast, short-lived eruption. The two eruptions seen in Figure \ref{fig:Aug2010} provides an example of this situation. 

The matches to generic eruption continues to be slightly over 50\%. In this larger time span, the two sets (filament eruptions and generic eruptions) are reasonably independent, with the latter catching a larger range of behavior which is left to the discretion of the annotator. The fact that there is not higher success rate here is to be expected, since EP reports only a single location for a given time sample. If multiple eruptions are occurring, all but one of them will be missed. This is another issue that  will be addressed in the characterization module. 

The overall accuracy of manual entries, finding less than a quarter of significant time periods identified by EP and with only 11\%  overlapping spatially is noteworthy. Their accuracy appears to be uncorrelated with the eruption's duration or magnitude. There may be some relation to spatial extent, but in reviewing a smaller sample with members of the annotation team, the main source of discrepancy is most likely a lack of attention or interest. Each annotator has particular interests related to their individual research, and there is a clear correlation between their success at this task and their interests.

  \begin{figure*}
   \centering
   \includegraphics[width=\columnwidth]{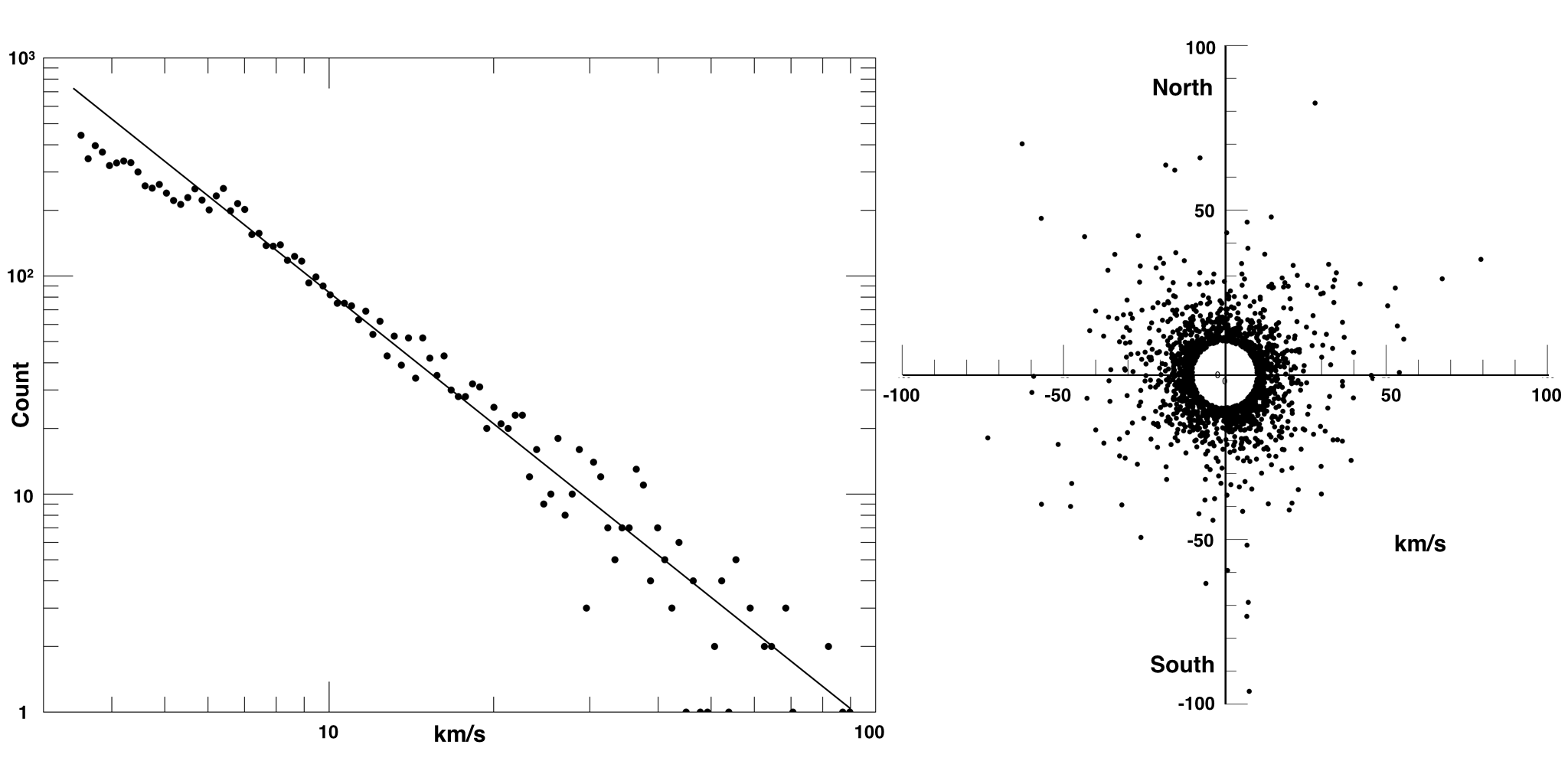}
   \caption{\small(Left) Histogram of speeds (black dots) of all eruptions from 5/15/2010 to 7/12/2014. Eruptions appear to possess a power law distribution near that of an inverse square (solid line). (Right) A polar plot of the velocity vectors show them to be reasonably isotropic with a maximum speed of 96.2 km/s. For clarity, only speeds over 10 km/s are displayed).} 
   \label{fig:speeds}
   \end{figure*}

\subsection{Statistical properties}
The EP module described above was run over the entire SDO mission up to July 12, 2014, thus spanning just over four years. Here we give an overview of the statistical properties of this sample. The left panel in Figure \ref{fig:speeds} displays the histogram of peak speeds detected for each recorded eruption. The distribution appears to have power law dependence on the speed, with the largest event having a speed of 96 km/s. These are consistent with previous studies, such as \cite{Gopal2003} and \cite{Wang}. The right panel in Figure \ref{fig:speeds} displays the distribution of velocities as a polar plot. There does not appear to be a significant directional bias in the sample. 

 \begin{figure*}
   \centering
   \includegraphics[width=\columnwidth]{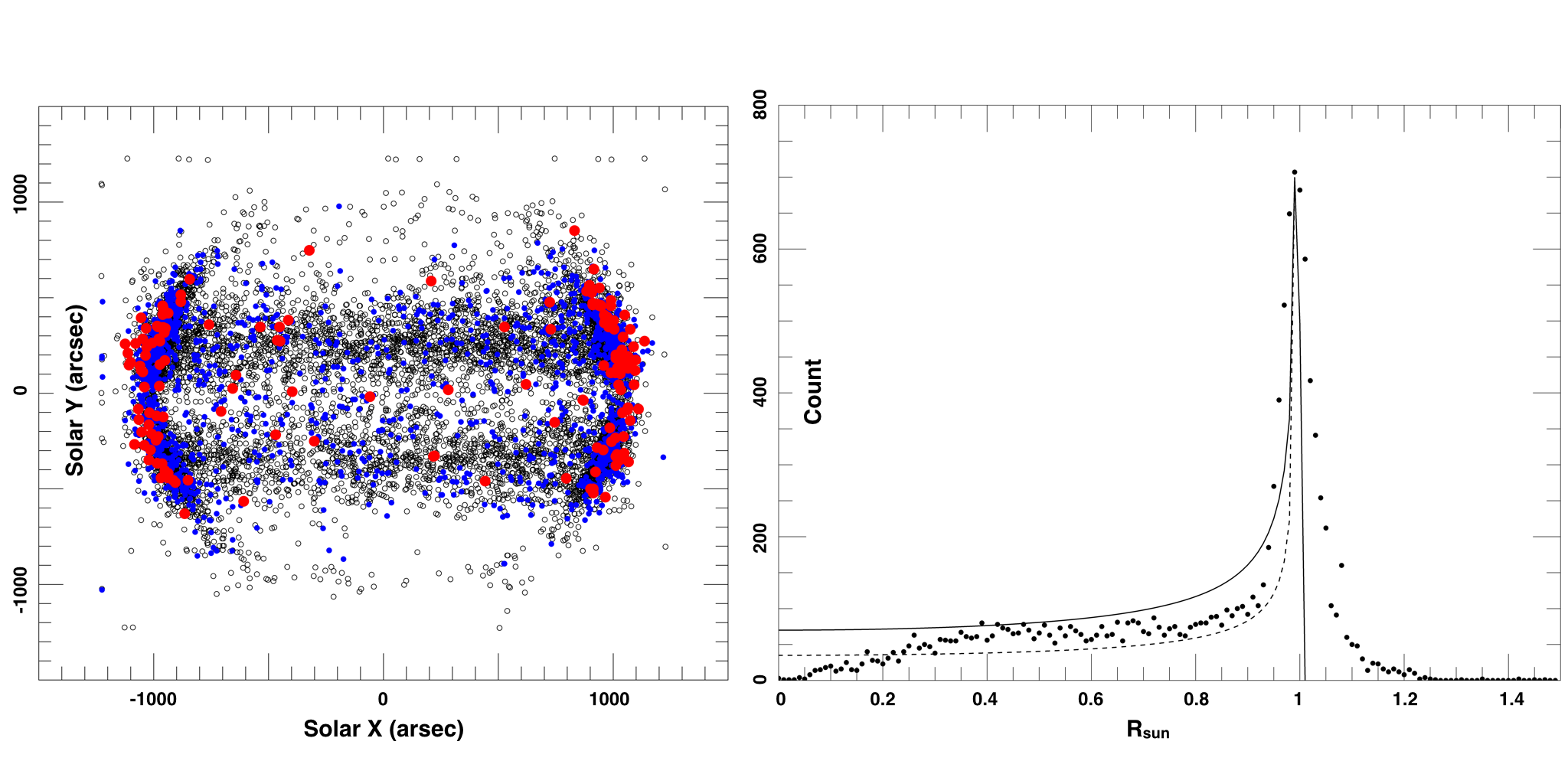}
   \caption{\small (Left) The distribution of all eruptions from 5/15/2010 to 7/12/2014 show clustering near activity belts and the limb. Black, blue and red circles display weak ($\langle10$ km/s), medium ($\langle30$ km/s) and strong ($\rangle30$ km/s) eruptions respectively. (Right) Histogram of radial positions. Dots represent number of events  in 10 arc second bins. The solid line is a theoretical distribution assuming the formation layer is a uniform 15 Mm thick, while the dashed is same for a 3.5 Mm layer.  Both lines are scaled to match the maximum. The distribution is compatible with a layer thickness somewhere within the range of 3-15 Mm.} 
   \label{fig:dist}
   \end{figure*}

Figure \ref{fig:dist} displays the spatial distribution of all events over this period. Eruptions are detected almost everywhere on the disk, as seen in the left panel, but are clearly clustered near the activity belts and the limb. This distribution appears to be independent of the magnitude of the events (as indicated by the color of the dots). There is a clear lack of eruptions reported near the poles which may be due to relatively slow-moving polar crown filament eruptions being masked by more dynamic regions as described in the last section.

The apparent clustering near the limb is examined in the right panel, where the histogram as a function of radius (r) for these events is displayed. The distribution rises from zero near disk center  ($r=0$) until the active region bands begin to contribute at $r \approx 0.4 R_{sun},$ where $R_{sun}$ is the solar radius. The distribution remains relatively constant between that point until near the limb ($r \approx R_{sun}$), where the number of eruptions climb rapidly before falling back to zero. This distribution is consistent with that expected in the case of a shallow, optically-thin, formation layer (between 3-15 Mm) containing a uniformly random distribution of eruptions. \cite{Gopal2003} report that (relatively large) eruptive prominences have heights between 1.1 to 1.5 $R_{sun}$. This suggests that there may be some scale dependence in the distribution that is neglected in our simple model, which might also explain the deviations from our  model for $r > R_{sun}$. Some level of scale dependence is expected in the structuring of the solar atmosphere by magnetic fields, as in the magnetic carpet model of \cite{Title}.

Another source of systematic error may result from projection effects. If all eruptions were predominantly radial, we would expect a radial dependence in the magnitude and direction of the reported velocities. \cite{Gopal2003} found this to be the case overall, but with many eruptions possessing tangential motions on the order of 10 km/s. In contrast, the motions we detect are randomly oriented and show no significant projection effects. We may resolve this discrepancy by noting that the former study was effectively tracking the centroid of a prominence while our method is measuring local velocities, which includes twisting, writhing and streaming motions that frequently accompany eruptions. Hence we can identify times and location of eruptions based on these associated motions regardless of where they appear on the disk. A more complete reckoning of these velocity contributions falls to the future characterization module.

\section{Conclusion}
We have developed an automated method for finding eruptions in the  lower solar atmosphere and have deployed it within the SDO/AIA Event Detection System which operates on the data as they arrive. The method has been found to measure velocities with statistical properties consistent with previous studies. The reported eruptions also appear to be consistent with those reported by human reviewers. The automated detections are  less prone to lapses in attention or skewed by personal interests, but may miss slow, long-duration eruptions.  They also provide a more complete characterization of eruptions by reporting both the location and plane-of-sky velocity. The reported events are found to be distributed in a layer near the solar surface and possess a power law distribution in peak speed. Details of these events, including summary movies, can be found using a variety of tools including Helioviewer\footnote{\url{http://helioviewer.org}}, iSolsearch\footnote{\url{http://www.lmsal.com/isolsearch}} and SolarSoft. As part of the HEK, they are automatically cross-referenced with solar datasets obtained by the Hinode (\cite{Kosugi}) and Interface Region Imaging Spectrograph (IRIS, \cite{BDP}) missions.

Subsequent papers will explore how these eruptions compare with those found with other automated processes recording in the HEK and will describe a characterization module that confirms and extracts more detailed information on the eruptions reported by the EP.

\begin{acknowledgements}

Thanks to Dr Paul Higgins for his helpful discussions and to Ryan Timmons and Sam Freeland for help implementing the module. This work has been supported by NASA under contract NNG04EA00C and Lockheed Martin Internal Research Funds. The editor thanks Jean-Fran\c cois Hochedez and two anonymous referees for their assistance in evaluating this paper.

\end{acknowledgements}



\end{document}